%% file: main.tex
\documentclass{article}
\usepackage{spconf,amsmath,graphicx}
\usepackage{amssymb}
\usepackage{multirow}
\usepackage{xcolor}
\usepackage{tabularx}
\usepackage{caption}
\usepackage{enumitem}
\usepackage{amsfonts,bm}
\usepackage{url}
\input{math_commands.tex}

\ninept
\title{End-to-end Multi-Person Audio/Visual Automatic Speech Recognition}
\name{Otavio Braga\textsuperscript{1}, Takaki Makino, Olivier Siohan, Hank Liao}
\address{Google Inc. \\ \textsuperscript{1}obraga@google.com}

\begin{document}
%
\maketitle
\begin{abstract}
Traditionally, audio-visual automatic speech recognition has been studied under the assumption that the speaking face on the visual signal is the face matching the audio. However, in a more realistic setting, when multiple faces are potentially on screen one needs to decide which face to feed to the A/V ASR system. The present work takes the recent progress of A/V ASR one step further and considers the scenario where multiple people are simultaneously on screen (multi-person A/V ASR). We propose a fully differentiable A/V ASR model that is able to handle multiple face tracks in a video. Instead of relying on two separate models for speaker face selection and audio-visual ASR on a single face track, we introduce an attention layer to the ASR encoder that is able to soft-select the appropriate face video track. Experiments carried out on an A/V system trained on over 30k hours of YouTube videos illustrate that the proposed approach can automatically select the proper face tracks with minor WER degradation compared to an oracle selection of the speaking face while still showing benefits of employing the visual signal instead of the audio alone.
\end{abstract}

\section{Introduction}
\label{sec:intro}
Audio-Visual Automatic Speech Recognition (A/V ASR) \cite{Makino19}\cite{Afouras_2018}\cite{Chung_2017} is a promising approach to robust ASR. Traditionally, it has been studied under ideal conditions where the face of the speaker is provided, while in a more realistic setting one has to decide at each point in time which face to use when multiple faces are on screen. In this work we address the impact of this problem on A/V ASR and propose a new framework that incorporates the face selection process directly into the A/V ASR model itself. This is a critical issue to practical A/V ASR systems and to the best of our knowledge this is the first work to directly address this problem.

Concretely, we can break down a conventional pipeline for A/V ASR into the following sequence of modules:
\begin{itemize}[leftmargin=*]
    \item {\it Face Tracking}: Which detects and tracks the faces in a video, and is in general conditions a solved problem in computer vision \cite{Schroff_2015}\cite{lugaresi2019mediapipe}.
    \item {\it Active Speaker Face Selection}: Given the face tracks and the audio, this module selects the speaking face and passes it along with the audio to the A/V ASR system \cite{Makino19}. At this step one would employ a separately trained model such as SyncNet or some of its newer incarnations \cite{ChungSyncNet_2017} \cite{ChungSyncnet2_2019}.

    \item {\it Audio-Visual Automatic Speech Recognition}: The A/V ASR system then operates on the audio and the face track picked by the active speaker selection system, outputting the predicted transcription for the video segment. Recent examples of A/V ASR architectures include \cite{Makino19}, \cite{Afouras_2018} and \cite{Chung_2017}.
\end{itemize}

Traditionally, the A/V ASR system is a separate model, trained with a single face track (assumed to be the speaking face) along with the audio track. This way, some level of robustness must be crafted into the system such that performance does not degrade when it is fed the wrong face track.

Our proposal is to discard the explicit active speaker selection system altogether and train a single model that can handle multiple face tracks. In addition to the audio track, instead of training the model with a single face track we train it with multiple tracks and learn how to gate the correct track to aid speech recognition with an attention mechanism. This is the main contribution of this paper.

Among the advantages of an end-to-end model like this we highlight the following:
\begin{itemize}[leftmargin=*]
    \item {\it Computational:} Each of the A/V active speaker and the ASR models need to rely on a separate visual frontend, 
    and each by themselves can easily dominate the total number of FLOPS of the entire model. However, low level visual features are typically transferable between computer vision tasks, so it seems redundant to have two separate sub-modules that are potentially playing similar roles.

    \item {\it Simplicity}: There is less coordination between subsystems with an end to end system. All we need is the output from a face tracker, which is a standard computer vision component these days \cite{lugaresi2019mediapipe}\cite{opencv_library}.

    \item {\it Robustness}: By not making an early hard-decision, our system is able to soft-select the active face track. Even when a high probability is assigned to the wrong face, the rest of the model is naturally allowed to adapt to that bad decision since the whole model is differentiable.
    
    On the other hand, when using a separate module for speaker selection one needs to deal with the dynamics of the discrete change of speaker over time, which can't be easily emulated, say, with an input augmentation strategy when training the ASR model.
\end{itemize}

In the following sections we detail our model, briefly describe the datasets used for training and evaluation, then present the results of our experiments.

Given the sensitive nature of the technology presented in this paper, it should be noted that this work abides by Google AI Principles \cite{GoogleAIPrinciples}\cite{Makino19}.

\section{Model}
\label{sec:model}

\subsection{Acoustic Features}
The 16kHz-sampled input audio is framed with 25ms windows smoothed with the Hann window function, with steps of 10ms between consecutive frames. We compute energies in 80 mel filter bank channels at each frame, compressing their range with a $\log$ function. We then fold every 3 consecutive feature vectors together, yielding a 240 dimensional feature vector every 30ms, which corresponds to acoustic features at about 33.3Hz. This last step is simply to reduce the number of unrolled LSTM steps further down in the ASR model encoder and speed up training. We denote the input acoustic features tensor by $\tA \in \R^{B\times T \times D_A}$, where $B$ is the batch size, $T$ is the number of time steps and $D_A$ ($=240$) the dimension of the acoustic features.

\subsection{Visual Frontend}

\subsubsection{Input Videos and Synchronization}
The videos in our training set have frame rates ranging from around 23 to 30 fps. In order to make the visual input uniform, we synchronize the videos with the acoustic features directly in the model input layer with nearest neighbor interpolation in the time dimension. I.e., given the acoustic features sample rate $a_\mathrm{fps} (\approx 33.3\mathrm{Hz}$ in our case) and the video frame rate $v_\mathrm{fps}$ (an attribute of each video), we can gather the closest video frame $i_v$ corresponding to each time step $i_a \in [1,\ldots,T_a]$ of the acoustic features:
\begin{equation}
i_v = \mathrm{round}\left(i_a  \frac{v_\mathrm{fps}}{a_\mathrm{fps}}\right).
\end{equation}

Previous experiments \cite{Makino19} have shown that simple nearest neighbor interpolation with a fixed frame rate does not degrade performance on A/V ASR when compared to more elaborate interpolation schemes, so we adopt this strategy for the sake of simplicity. Unlike \cite{Makino19}, instead of computing the acoustic features at a variable sample rate synchronized to the video frame rate here we conduct all our experiments with the videos synchronized to the sample rate of the acoustic features.

In the spatial dimension, we crop the full face tracks around the mouth region to generate $128\times128$ size images, with RGB channels normalized between $-1$ and $1$.

\subsubsection{3D ConvNet Layers for Visual Features}
For the actual trainable model, on top of the synchronized video we compute visual features $\tV \in \R^{B\times T\times D_v}$ with a 3D ConvNet \cite{Lecun98}. For the exact parameters we use a ``thick'' VGG-inspired \cite{Simonyan15} setup, as in \cite{Makino19}, with a stack of 5 layers of 3D convolutions, with $3\times3\times3$ kernels and consecutive channels of sizes $[3, 64, 128, 256, 512, 512]$ (i.e., 3 input and 64 output channels on the first layer, 64 input and 128 output channels for the second layer, and so on). We use a stride of 1 in all dimensions, except on the first layer, where we use a stride of 2 in the spatial dimensions, with SAME padding in the time dimension and VALID padding in the spatial dimensions. ReLU activation functions follow the convolutions, except on the last layer, and then $2\times2$ spatial max pooling operations in the spatial dimension, except the fourth layer. Therefore, the output layer has the same number of time steps as the input, and at each step we have a $1\times1\times 512$ tensor, or, simply, a $D_v = 512$ dimensional feature vector. In order to speed up training, we also employ group normalization \cite{wu2018group} in the spatial dimension with 32 groups after each activation function.

\begin{figure}[!t]
\begin{minipage}[b]{.48\linewidth}
  \centering
  \centerline{\includegraphics[width=4.0cm]{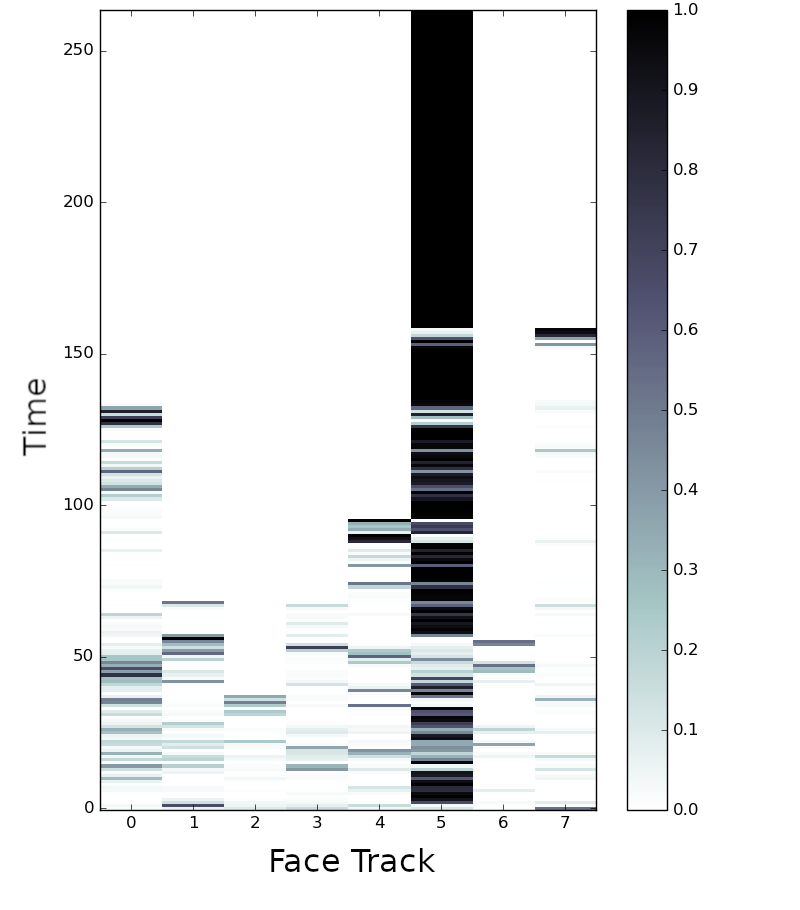}}
\end{minipage}
\hfill
\begin{minipage}[b]{0.48\linewidth}
  \centering
  \centerline{\includegraphics[width=4.0cm]{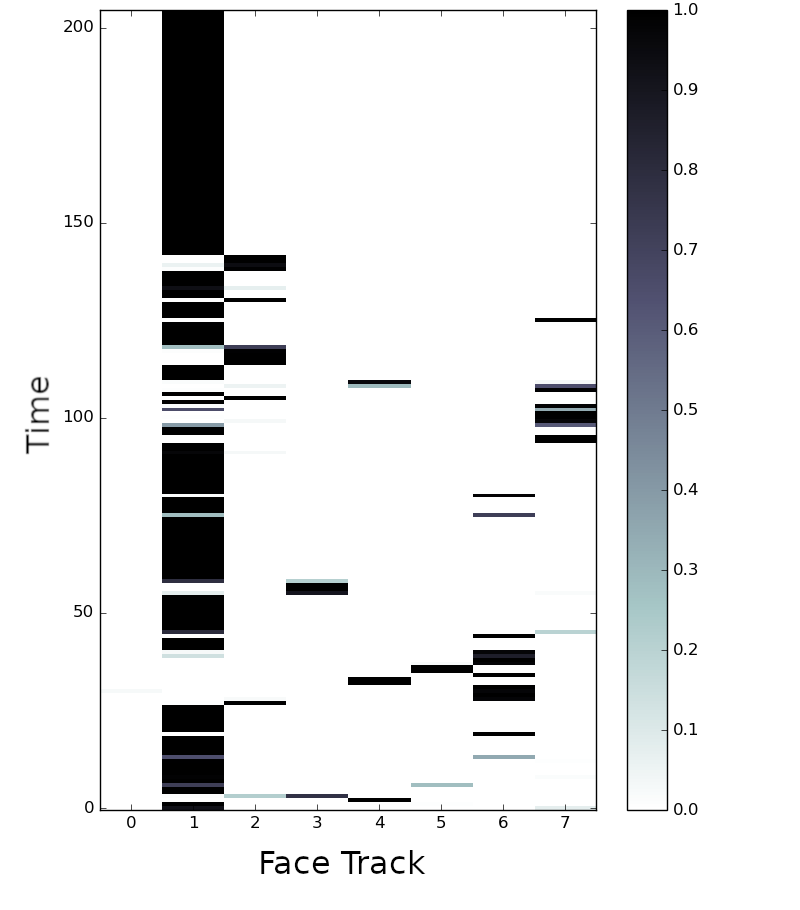}}
\end{minipage}
\caption{Examples of attention weights $\alpha_{b,:,:}$ among competing video tracks for a minibatch element $b$. Horizontal axis is the index of the competing video tracks, i.e., rows sum to one. See Section~\ref{sec:attention}.}
\label{fig:attention}
\end{figure}

\subsection{Batch Gating Attention}
\label{sec:attention}
We treat slices $\tV_{b,:,:} \in \R^{T\times D_v}$ of the visual features tensor along the batch dimension as visual features of competing faces and employ an attention module in order to soft-select the correct one.

Given two tensors $\tQ \in \R^{B\times T\times D_q}$ and $\tV \in \R^{B\times T\times D_v}$, for each minibatch element $b$ and time step $t$ we compute a new feature tensor
$\tV'$ by ``gating'' its batch dimension at each time step. Each vector
$\tQ_{b,t,:} \in \R^{D_q}$ computed from the acoustic features plays the role of a query to an attention block on $\tV_{:,t,:} \in \R^{B \times D_v}$ along the batch dimension.  In the next section we describe exactly how we compute the query tensor $\tQ$.

Concretely, for the attention score we use a bilinear function with parameter matrix
$\tW \in \R^{D_q \times D_v}$. In Einstein summation notation, the score is given by
\begin{equation}
S_{ijk} = Q_{ijl}(\tW\tV_{k,j,:})_l, \quad \textrm{with} \quad \tS \in \R^{B\times T\times B}.
\end{equation}
Note that the product $\tW\tV_{k,j,:} \in \R^{D_q}$ is a simple matrix-vector product used to project the visual features $\tV_{k,j,:} \in\R^{D_v}$ to the same last dimension as the query tensor $\tQ$ , implemented as a 1D convolution of size $1$ (which is then used to compute the attention score with a dot product). Other attention score functions (see \cite{BahdanauCB14}\cite{Luong_2015}) can evidently be used here.

The attention scores are normalized {\it along the batch dimension of the visual features} with a softmax in order to produce the attention weights:
\begin{equation}
\label{eq:softmax}
\alpha_{btk} = \frac{e^{S_{btk}}}{\sum_{l} e^{S_{btl}}}, \quad \textrm{with} \quad \alpha \in \R^{B\times T \times B},
\end{equation}
and the attention weighted feature tensor $\tV'$ then becomes
\begin{equation}
V'_{btk} = \alpha_{bti} V_{itk}, \quad \textrm{with} \quad \tV' \in \R^{B \times T \times D_v}.
\end{equation}

Therefore, we end up with a new tensor of visual features, but now soft-selected among the competing visual feature vectors on the minibatch. For training, this is an easy and convenient extension to the frontend since training is already done in minibatches and we take advantage of that in order to train a model supporting multiple speakers. Note that since we use distributed training, by the batch mixing dimension we mean the batch in a single core (8 for our experiments), and not the effective combined batch size (say, $1024 = 8\times128$ when training on 128 cores).

At inference time, the visual features are simply the visual features of all the face tracks on the input video. This way, we can naturally handle an arbitrary number of face tracks, with the additional interpretability of the attention weights $\alpha$
giving us a measure for the speaking probability of each face track in time. Figure \ref{fig:attention} shows two examples of the attention weights in time.

\subsubsection{1D ConvNet for Attention Query Layer}

We use a 1D ConvNet in order to compute the attention query vector $\tQ$ for the attention module described above. We roughly double the receptive field of the visual frontend by using 5 layers of 1D convolutions with kernel sizes of 5 (just time dimension here) and SAME paddings for all layers. This way, each output time step corresponds to a receptive field of 25 steps in the input, or 750ms since we are computing acoustic features every 30ms. We use [240, 256, 256, 256, 512, 512] channels, with 240 input channels for the acoustic features, ReLU activation functions and batch normalization between the layers. We do not use any pooling layers.

\begin{table}[!t]
\begin{center}
\begin{tabularx}{\linewidth}{XXccc}
\multicolumn{1}{l}{\bf Test Set} & \multicolumn{1}{l}{\bf Added Noise} & \multicolumn{1}{c}{\bf A} & \multicolumn{1}{c}{\bf A+V} \\
\hline \\ [-1.5ex]
\multirow{5}{*}{YTDEV18} & \multicolumn{1}c{--} & 17.7 & 16.4 \\
                         & Babble, 20dB \qquad  & 18.6 & 16.9 \\
                         & Babble, 10dB         & 24.1 & 20.4 \\
                         & Babble, 0dB          & 62.8 & 53.4 \\
                         & Overlap              & 37.1 & 33.9 \\ [1.0ex]
\hline \\ [-1.5ex]
LRS3-TED                 & \multicolumn{1}c{--} & 4.3  & 4.3 \\
\end{tabularx}
\end{center}
\caption{Baselines: WER ($\%$) for audio only ({\bf A}) and audio-visual ({\bf A+V}) ASR models trained with a single speaker with clean speech.}
\label{table:baselines}
\end{table}

\subsection{Audio-Visual RNN-T Model}

With the input acoustic features $\tA \in \R^{B\times T \times D_A}$ and the attention-weighted visual features $\tV' \in \R^{B\times T \times D_V}$ in hands, we then concatenate both tensors along the last dimension to yield the combined feature tensor $\tF = [\tA; \tV'] \in \R^{B\times T \times (D_A + D_V)}$, which is then fed into our original A/V ASR model (or any audio-visual model for that matter).

For ASR, we employ a standard RNN-T model \cite{graves2012sequence}, \cite{graves2013speech} , with a stack of 5 BiLSTM of 512 units on each direction using layer normalization for the encoder, and 2 LSTM layers of 2048 units with character tokens for the decoder.

\section{Datasets}
\label{sec:datasets}

\subsection{Single Speaker Training and Evaluation Datasets}

For training, we use 37k hours of transcribed short YouTube video segments extracted with the same semi-supervised procedure described in \cite{Shillingford_2019}\cite{Makino19}. We extract short segments where the force-aligned user uploaded transcription matches the transcriptions from a production quality ASR system. From these segments we then keep the ones in which the face tracks match the audio with high confidence. See \cite{Shillingford_2019} and \cite{Makino19} for more details of the pipeline.
 
For evaluation, we rely on base datasets:
\begin{itemize}[leftmargin=*]
    \item {\it YTDEV18} \cite{Makino19}: This dataset is composed of 25 hours of manually transcribed YouTube videos, not overlapping with the training dataset. We track the faces on screen and pick the segments where the audio and video tracks match with the same procedure used to extract the training data. Therefore, by design, the faces extracted from the video correspond with high probability to the speaker in the audio. The result is a dataset with around 20k utterances.
    
    In order to better evaluate the impact of the visual modality, we also evaluate on noisy conditions by adding babble noise randomly selected from the NoiseX dataset \cite{Varga1993AssessmentFA} at 0dB, 10dB and 20dB to each utterance. Moreover, we evaluate the effect of overlapping speech by randomly adding another utterance from the same dataset at the same level to the beginning and end of each utterance.
    
    \item {\it LRS3-TED Talks} \cite{Afouras18d}: For completeness, we also evaluate on this dataset since it is the largest publicly available dataset for A/V ASR. Again, it contains only pairs of audio and single face tracks that match the audio. However, we believe this dataset is not challenging enough as the performance quickly saturates with an audio only system and we don't observe significant gains when adding video. Videos are recorded by professionals, and the quality of both the audio and video are typically above the quality of a general video one would encounter in a realistic setting.
\end{itemize}

\subsection{Datasets with Multiple Face Tracks}
\label{sec:ms-datasets}
In order to evaluate our model in the scenario where multiple face tracks are simultaneously visible in a video, we construct a new evaluation dataset as follows: On the single track evaluation sets described in the previous section, at time $t$ both the acoustic and visual features from the corresponding face are available. To build a dataset with $N$ parallel face tracks we start from the single track set, and for every pair of matched audio and face video track we randomly pick other $N-1$ face tracks from the same dataset. Therefore, during evaluation at each time step we have the acoustic features computed from the audio and $N$ candidate visual features, without knowing which one matches the audio. We generate separate datasets for $N = 1, 2, 4, 8$.

Note that these are potentially harder datasets than what we would normally encounter in practice since all faces are always speaking here, while it is an easier task to differentiate between a speaking and a non-speaking face video as we would encounter most of the time in a meeting scenario, for example. We adopt this evaluation protocol nonetheless and leave as future work to evaluate our model on more realistic datasets.

\begin{table}[!t]
\begin{center}
\begin{tabularx}{\linewidth}{XXcccc}
\multirow{3}{*}{\bf Test Set} & \multirow{3}{*}{\bf Added Noise} & \multicolumn{4}{c}{\bf Face Tracks} \\
                              &                                  & {\bf 1} & {\bf 2} & {\bf 4} & {\bf 8} \\
\hline \\ [-1.5ex]
\multirow{5}{*}{YTDEV18} & \multicolumn{1}c{--} & 16.4 & 16.5 & 16.6 & 16.7 \\
                         & Babble, 20dB         & 16.9 & 17.0 & 17.1 & 17.3 \\
                         & Babble, 10dB         & 20.4 & 20.6 & 21.0 & 21.3 \\
                         & Babble, 0dB          & 53.4 & 55.3 & 57.1 & 58.9 \\
                         & Overlap              & 33.9 & 34.3 & 35.1 & 35.5 \\ [1.0ex]
\hline \\ [-1.5ex]
LRS3-TED                 & \multicolumn{1}c{--} & 4.3 & 4.3 & 4.3 & 4.4 \\
\end{tabularx}
\end{center}
\vspace{-1mm}
\caption{Evaluation (WER) of multi-person model on datasets created with an increasing number of simultaneously speaking face video tracks while using only one of the audio tracks.}
\label{table:wer-vs-num-speakers}
\end{table}

\section{Experiments}
\label{sec:experiments}

\subsection{Baselines}
\label{sec:baselines}
Our baselines are two RNN-T models: an audio only and an audio-visual model, the latter trained with pairs of matching audio and a single face track. The audio only model is identical to the A/V model, excluding the video input and 3D ConvNet layers. Table~\ref{table:baselines} shows the WER on our evaluation sets for the two models. The WER on the audio-only model gives an upper bound for the error we can incur while still showing benefits of using the visual signal, and the A/V WER a lower bound on the error we can hope to achieve when evaluating on multiple face tracks. On the YTDEV18 set we observe from about $7\%$ to $15\%$ relative WER reduction between the audio-only and the A/V model.

\begin{figure}[!t]
\begin{minipage}[b]{\linewidth}
  \centering
  \centerline{\includegraphics[width=0.8\textwidth]{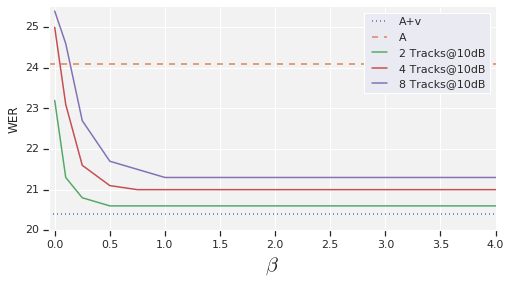}}
\end{minipage}
\vspace{-5mm}
\caption{Changing the softmax temperature: For $\beta = 0$, the model uniformly selects the visual features. As $\beta \to \infty$, it gracefully converges to picking the visual features of the track with maximum attention weight. See Section~\ref{sec:low-entropy} for the details.}
\label{fig:entropy}
\end{figure}

We train using the Adam optimizer with parameters $\beta_1=0.9$ and $\beta_2=0.98$, and a learning rate schedule broken down into 3 stages: First linearly increasing the learning rate up to 0.001 until 32k steps, then keeping it constant until 60k steps, then exponentially decaying it until 200k steps. At training we restrict our source and target sequences to a maximum length of 360 and 512, respectively.

\subsection{Evaluation with an Increasing Number of Face Tracks}

Next, we evaluate the multi-person model with an increasing number of parallel video tracks, which simulates the type of video recordings one would find in a meeting transcription application with multiple people on screen, for example. We build separate evaluation sets using the procedure described in Section~\ref{sec:ms-datasets}.

We train the multi-person model by initializing the shared weights from the baseline A/V ASR model, which is a convenient characteristic of our extension since it significantly speeds up training. The end-to-end model is again trained with RNN-T loss.

Results are shown in Table~\ref{table:wer-vs-num-speakers} for different numbers of parallel tracks and noise levels. While there is some degradation in performance as we add more speakers, the performance still remains consistently better than the audio only model, as in the single track baseline models evaluated on Table~\ref{table:baselines}, indicating that the visual signal is still helpful. For example, even with 8 speakers with noise at 10dB we still observe a $12\%$ relative WER improvement to the audio-only model, and a $4\%$ degradation of the audio-visual model with oracle face selection. Note, however, that we only trained with the original audio, so by extrapolating the results reported in \cite{Makino19} we believe that including explicitly corrupted speech in the training would stretch the benefits even further. We leave this evaluation for follow-up work though.

Finally, note that our setup permits training just the attention layer connected to a frozen pretrained A/V ASR model, including freezing the 3D convolutions. We found that this strategy leads to slightly worse performance as compared to training the whole model, though the results are still better than an audio only model.

\subsection{Varying the Entropy}
\label{sec:low-entropy}

Our model soft-selects a speaking face among all parallel face tracks, and we can gracefully transition from a completely random to a hard selection if we introduce an ``inverse temperature'' parameter $\beta$ into the softmax in equation \ref{eq:softmax}. At inference time, in order to evaluate our model as we blur or sharpen the probability distribution around the maximum we can evaluate $\alpha_{btk}^{(\beta)} = e^{\beta S_{btk}}/\sum_{l} e^{\beta S_{btl}} \in \R^{B\times T \times B}.$

As $\beta\rightarrow\infty$, $\alpha^{(\beta)}_{b,t,:}$ converges to a one-hot vector with 1 at index $k = \textrm{argmax}_{k'}S_{ijk'}$, effectively making our soft-selection converge to a hard decision rule of picking the track with the maximum attention score. On the other hand, as $\beta \to 0$ our model converges to a uniform distribution among the face tracks. Figure~\ref{fig:entropy} shows the effect of $\beta$ on the WER for the datasets with different number of competing speakers. Note that the learned model ($\beta = 1$) already has low entropy to start with, so when we push the limit to make a hard selection with the face track with max probability ($\beta = \infty$) the WER stays about the same.

\subsection{Training the Attention Module with Cross Entropy}
\label{sec:ce-loss}
Lastly, we compare the performance of the end-to-end model trained with RNN-T loss with an equivalent model where the face track selection layer is directly trained with cross entropy loss.

Specifically, for each timestep $t$ and minibatch element $b$, given one example of acoustic features within the minibatch, we are trying to guess the corresponding visual features of the correct face track among all face tracks within the minibatch, $\tV_{:,t,:}$. We know by construction that $\tA_{b,:,:}$ corresponds to $\tV_{b,:,:}$, so if we denote by $p_{btk}$ the ground truth probability of acoustic features of minibatch element $b$ at time $t$ corresponding to the visual features of minibatch element $k$, we have $p_{btk} = [b=k]$ and the cross entropy loss is thus:
\begin{equation}
\label{eq:CE}
L = \frac{1}{BT}\sum_{b=1}^B\sum_{t=1}^T\sum_{k=1}^B-[b = k]\log \alpha_{btk}.
\end{equation}

Here we freeze the 3D convolutions borrowed from the pretrained A/V ASR model and only train the rest of the attention layer so that we can share the 3d ConvNet between the speaking face selection and the A/V ASR modules.

The classifier is trained until convergence, and results are in Table~\ref{table:ce-loss-wer-vs-num-speakers}. We can observe a slight decrease in performance in some of the datasets as compared to the end-to-end model, but still always better than the audio only model. We leave for a future work a more thorough comparison between the two modes of training, including combining the two losses in a multi-task learning setup \cite{Caruana1993MultitaskLA}.

\begin{table}[!t]
\begin{center}
\begin{tabularx}{\linewidth}{XXcccc}
\multirow{3}{*}{\bf Test Set} & \multirow{3}{*}{\bf Added Noise} & \multicolumn{4}{c}{\bf Face Tracks} \\
                              &                                  & {\bf 1} & {\bf 2} & {\bf 4} & {\bf 8} \\
\hline  \\ [-1.5ex]
\multirow{5}{*}{YTDEV18} & \multicolumn{1}c{--} & 16.4 & 16.8 & 17.3 & 17.6 \\
                         & Babble, 20dB         & 16.9 & 17.3 & 17.7 & 18.2 \\
                         & Babble, 10dB         & 20.4 & 21.3 & 22.1 & 23.0 \\
                         & Babble, 0dB          & 53.4 & 56.9 & 59.8 & 62.3  \\
                         & Overlap              & 33.9 & 35.3 & 36.3 & 36.9  \\
\end{tabularx}
\end{center}
\vspace{-2mm}
\caption{Evaluation (WER) with attention module trained explicitly with cross entropy loss (See Section~\ref{sec:ce-loss}).}
\label{table:ce-loss-wer-vs-num-speakers}
\end{table}

\vspace{-1mm}
\section{Conclusions}
\label{sec:conclusions}

We have presented a model that is able to handle end-to-end the general case of A/V ASR where multiple speakers are simultaneously visible on screen. Despite being a typical scenario for practical systems, we believe that it has been overlooked in the A/V ASR literature so far.\looseness=-1

In the spirit of differentiable programming, we introduced an attention mechanism to couple a face selection model with an audio-visual speech recognition system. Instead of manually crafting rules on top of a speaker selection module, we let gradient descent do the programming for us. We trained our system with over 30k hours of video segments, and showed that it is able to handle multiple face tracks with minor degradation while still showing improvements over an equivalent model trained with the acoustic signal alone.

From here, we would like to evaluate our model in more dynamic scenarios where speakers are taking turns, and expand the comparison with explicitly trained speaking face selection modules. Also, drawing from the positive results on single person A/V ASR, we plan to train our multi-person model with corrupted speech in the hope that the benefits of the visual signal will stretch even further.

Lastly, putting the specific architecture our model aside, we believe this is a promising general framework for multi-person audio-visual speech recognition models and we hope to push this research direction further.

\vfill\pagebreak

\vfill\pagebreak

\bibliographystyle{IEEEbib}
\bibliography{strings,refs}

\end{document}

%% file: math_commands.tex
\def\eqref#1{equation~\ref{#1}}

\def\1{\bm{1}}

\DeclareMathAlphabet{\mathsfit}{\encodingdefault}{\sfdefault}{m}{sl}
\SetMathAlphabet{\mathsfit}{bold}{\encodingdefault}{\sfdefault}{bx}{n}
\newcommand{\tens}[1]{\bm{\mathsfit{#1}}}
\def\tA{{\tens{A}}}

\def\tF{{\tens{F}}}

\def\tQ{{\tens{Q}}}

\def\tS{{\tens{S}}}

\def\tV{{\tens{V}}}
\def\tW{{\tens{W}}}

\newcommand{\R}{\mathbb{R}}

